\documentclass[preprint,showpacs,amsmath,amssymb]{revtex4}
\usepackage{amsmath,amssymb}
\usepackage{graphicx}
\usepackage{float}

\begin{document}

\title{Connecting Structural Relaxation with the Low Frequency Modes in a Hard-Sphere Colloidal Glass}
\author{Antina Ghosh$^{1*\dagger}$,
Vijayakumar Chikkadi$^1$\footnote{equally contributing authors}\footnote{corresponding authors :antinag@gmail.com,vijayck07@gmail.com},
Peter Schall$^1$ and Daniel Bonn$^{1,2}$.}

\affiliation{
$^{1}$Institute of Physics, University of Amsterdam, Science Park 904, 1098 XH Amsterdam, The Netherlands\\
$^{2}$LPS de l'ENS, 24 Rue Lhomond, 75321 Paris cedex 05, France}

\begin{abstract}
Structural relaxation in hard-sphere colloidal glasses has been studied using confocal microscopy. The motion of individual particles is followed over long time scales to detect the rearranging regions in the system. We have used normal mode analysis to understand the origin of the rearranging regions. The low frequency modes, obtained over short time scales, show strong spatial correlation with the rearrangements that happen on long time scales.
\end{abstract}
\maketitle
Supercooled molecular liquids, polymers or metallic systems, when cooled or compressed in a way so as to prevent crystallization, undergo a glass transition. The transition to the glassy state is characterized by the slowing down of the dynamics of the system accompanied by a spectacular increase in the viscosity, to values on the order of $\sim10^{13} Pa~s$ \cite{angell}. Both features are usually attributed to the fact that the motion of individual molecules or particles become arrested due to the presence of the cage formed by neighboring particles. However, on longer timescales the particles escape from their cages, leading to the relaxation of system.

In recent experiments on colloidal glasses, the motion of individual particles was followed using confocal microscopy to investigate such cage rearrangement events in quiescent \cite{c_glass,glass_weeks} as well as sheared \cite{Schall,shear_weeks} systems. It was found that the particle dynamics is very heterogeneous in the sense that some regions exhibited much stronger activity than others: some small parts of the system are more susceptible to rearrangements than others. Earlier studies have reported such heterogeneous dynamics in glass forming liquids and polymers \cite{ediger}.

Perhaps one of the main questions about glassy dynamics is then how to understand and eventually predict the relaxation events that involve rearrangement of molecules or particles. One clue comes from recent computer simulation studies of supercooled liquids, which suggest that more susceptible regions result from localized `soft' (i.e., low frequency) modes of the system \cite{Harrowell}. The soft modes in glassy systems is a subject of much interest lately \cite{Wyart}-\cite{ghosh-P}, as they are related to the anomalous low temperature properties of glasses. Using normal mode analysis, their existence has been reported recently in a few experiments on colloidal glasses \cite{ghosh} -\cite{Kaya}. However, there is very little experimental evidence that connect these modes with the relaxation events in colloidal glasses \cite{Kechen1}.

In this Letter, we identify the low frequency modes in a hard-sphere colloidal glass while simultaneously studying the rearrangements. This is done by following the motion of fluorescent colloids for a long time using rapid confocal microscopy. On the one hand, the particle motions can be analyzed in terms of the normal modes of the system. On the other hand, rearrangements can be identified by looking at the changes in neighbors. The main conclusion is that, indeed, the rearrangements happen along the softest modes of the system. We thus establish the structural origin of the rearranging regions by comparing them with the spatial maps of the amplitudes of the soft modes.

We study the motion of particles in a dense hard sphere colloidal glass using confocal microscopy. It is prepared by suspending sterically stabilized fluorescent polymethylmethacrylate (PMMA) particles in a density and refractive index matching mixture of Cycloheptyl Bromide and Cis-Decalin. The particles have a diameter of $\sigma = 1.3 \mu m$, with a polydispersity of $5\%$ to prevent crystallization. The organic salt TBAB (tetrabutylammoniumbromide) is used to screen any possible residual charges. A dense suspension of volume fraction $\phi\sim0.59$ was prepared by diluting a sediment that was centrifuged to random close packing ($\phi_{RCP}\sim0.64$). Since the particles are poly-disperse, the volume fraction of the sample is subject to an error of $\sim 2$\% \cite{hermes_dijkstra}.

\begin{figure}[h]
\begin{center}
\begin{minipage}{8.0cm}
\includegraphics[width=0.95\textwidth]{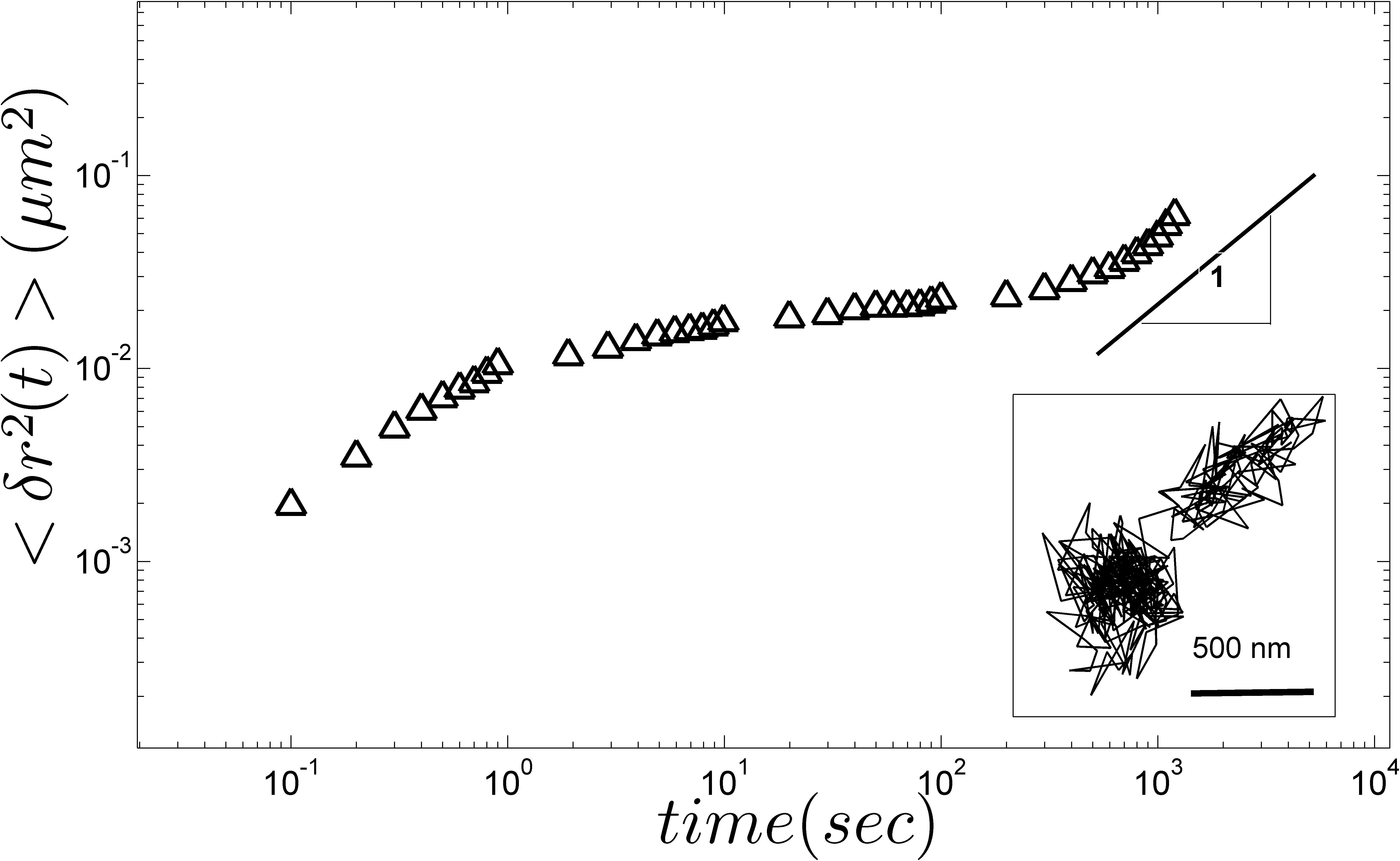}
\end{minipage}
\end{center}
\caption{Mean square displacement (MSD) of the particles is shown as a function of time over the whole period of measurement $\sim 1200s$. The dark line has a slope of unity. Inset shows a typical trajectory a particle that undergoes cage rearrangement where average position shifts from one location to another.}
\label{fig_MSD}
\end{figure}

Two dimensional slices through the 3d glass of fluorescent particles are acquired at a scanning speed of $10$ frames per second using a fast confocal microscope ({\it Zeiss LSM 5 live}) in a field of view of $100 \mu m \times 100 \mu m$. The image plane was chosen so that it is sufficiently $\sim 25 \mu m$ away from the coverslip to avoid any possible boundary induced effects. Images are acquired for a period of total $1200s$ which results in $12000$ independent snapshots. The positions of the particles are then identified in all the frames using standard particle tracking softwares and linked to construct the two dimensional trajectories. The mean square displacement (MSD) $<\delta r^{2}>$ of the particles computed using these trajectories is shown in Fig.~\ref{fig_MSD}. On short time scales, the rise of the average displacements is due to diffusion of particles before they start feeling the presence of their neighbors. Beyond this time scale, we observe a plateau in the MSD where the motion of a particle is restricted by a shell of nearest neighbors constituting the `cage'. It extends roughly up to $\sim 500s$ in the present experiments. Note that this times scale compares well with the previous measurements done on hard sphere colloidal glasses \cite{glass_weeks}. On even longer timescales, at the end of the plateau, the mean square displacement increases again. This corresponds to a long-time diffusive behavior known as the $\alpha$-relaxation \cite{pusey_megen,SRWilliams} regime and is usually attributed to cage rearrangements. The average displacement in the long-time diffusive regime is linear in time: $<\delta r^{2}> \sim t$, as follows from Fig.~\ref{fig_MSD}: the data in this region is indeed parallel to a line of slope unity. The effect of aging over the time scale of our measurement was found to be negligible.

However what is not clear from these averaged quantities is the possible heterogeneity of the dynamics, as suggested by previous experiments \cite{c_glass}. The inset to Fig.~\ref{fig_MSD} shows a single particle trajectory; there is typical 'cage rattling motion' followed for this particle by a rearrangement, in which the average of the center of mass changes positions. This does not happen for all particles over the observation time, and such events can be used to see where rearrangements happen.  To investigate this, we compute the square of the {\it relative} displacements $\it{d_{i}^{2}} (\Delta t)$ of each particle with respect to its nearest neighbors over a time interval $\Delta t$ from the reconstructed trajectories $x_{i}(t), y_{i}(t)$. The relative displacement $\it{d_{i}^{2}}$ is defined as :
%%%%%%%%%%%%%%%%%%%%%%%%%%%%%%%%%%
\begin{eqnarray}
d^{2}_{i}(\Delta t) = \frac{1}{n}\sum_{j=1}^{n}\left|\mathbf{\Delta r}_{ij}(t +\Delta t)-\mathbf{\Delta r}_{ij}(t)\right|^2 , \label{eq_D}
\end{eqnarray}
%%%%%%%%%%%%%%%%%%%%%%%%%%%%%%%%%%
where $n$ denotes the number of nearest neighbors that is determined using the method of Delaunay triangulation and $\mathbf{\Delta r}_{ij}(t) = \mathbf{r}_i-\mathbf{r}_j$ denotes the difference vector of the particles {\it i,j} at time $t$; this expression of $\it{d_{i}^2}$ is in fact motivated by the non-affine measure of plasticity used in sheared amorphous solids \cite{Falk}.

\begin{figure}[h!]
\begin{center}
$\begin{array}{c}
\begin{minipage}{5.2cm}
\includegraphics[width=0.95\textwidth]{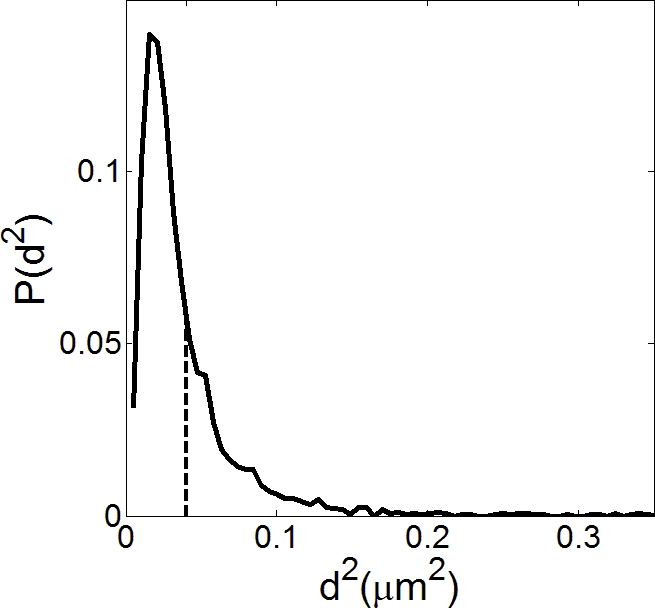}
\end{minipage}
\begin{minipage}{4.9cm}
\includegraphics[width=0.95\textwidth]{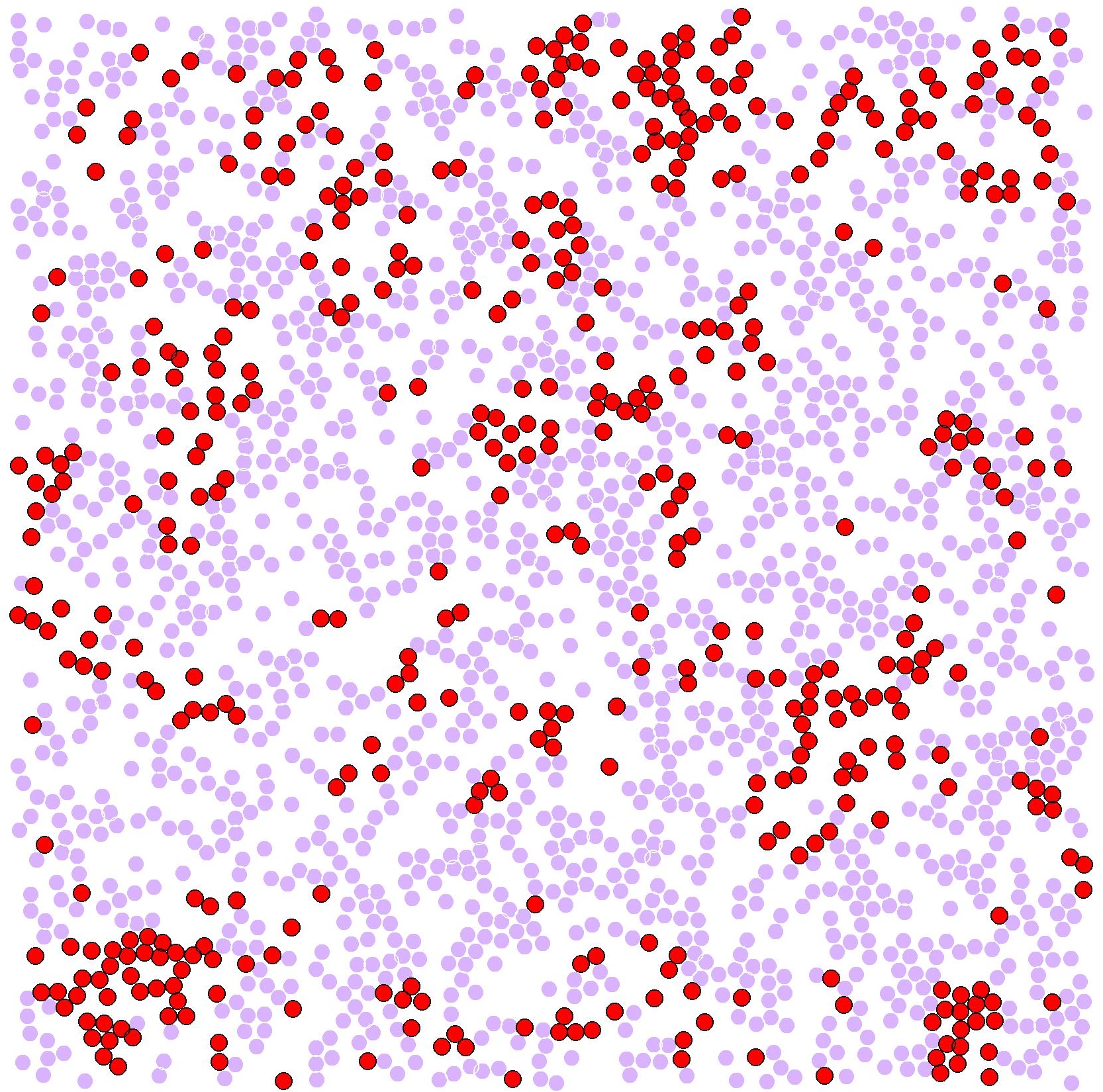}
\end{minipage}
\begin{minipage}{4.9cm}
\includegraphics[width=0.95\textwidth]{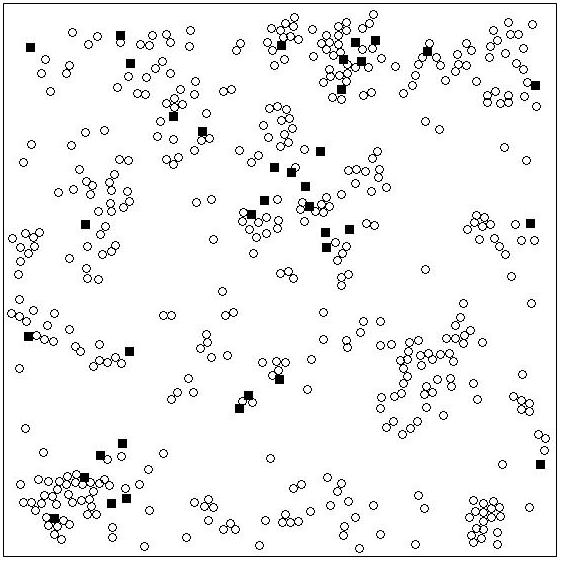}
\end{minipage}
\end{array}$
\end{center}
\caption{Left: The probability distribution function of the relative displacements $d_{i}$ for the present suspension
measured over a time interval $\Delta t = 1200 s$. The mean of this distribution is $<d_{i}^2>=0.039\mu m^2$, and it indicated by the dashed vertical line.
Center: A spatial map of the particles where the highly mobile particle with a $d_{i}^2 > r_{c}^2 (0.06\mu m^2)$
are shown by the red circles along with the less active particles denoted by pink circles.
Right: Superposition of the particles that have relative displacement $d_i^2 > r_c^2$ (circles) and the particles that have lost three or more neighbors (squares). Almost all the particles losing neighbors belong to the regions of larger relative displacement.}
\label{pdf}
\end{figure}

Figure~\ref{pdf}(a) shows the probability distribution function of the relative displacements of the particles evaluated over a time scale of $\Delta t \sim 1200s$. The extended tail of the distribution implies the presence of large displacements in the system. To detect the highly mobile particles, we tag all those particles that undergo a relative displacement beyond a cutoff value $r_{c}^2$. This cutoff $r_{c}=0.245 \mu m$ is $\sim 1.25$ times the square-root of the mean of the above distribution. The spatial distribution of the particles that have ${d_{i}^2}>r_c^2$
is shown in Fig.~\ref{pdf}(b). Here, the red circles refers to the particles with large $ d^{2}_{i} (> r_{c}^2)$ in a background of less active particles shown in pink. The particle with large $d_{i}^2$ appear in the form of clusters in the field of view. Our observation shows that particles in these regions undergo significant change in their neighborhood.

To demonstrate this, we identify the particles that change neighbors. The neighbors of each particle at any instant of time $t$ can be identified using the method of Delaunay triangulation. We then compare the neighbors of each particle at  time $t$ and $t+\Delta t$. This gives us directly the number of neighbors lost or changed around a particle over a time interval $\Delta t$. In simulations of supercooled liquids \cite{Harrowell}, the irreversible rearrangements were identified as particles that have lost four neighbors. However, in experiments there are very few particles that lose four neighbors over the time scale of our observation. So, we identify those particles that have lost three or more neighbors in our data. Figure~\ref{pdf}(c) shows a plot of these particles (squares) along with the clusters of highly mobile particles. We observe that the majority of the particles that have lost three or more neighbors belong to the regions of large relative displacements. This establishes that $d_{i}^2$ is a good way to detect rearranging regions in the system.

\begin{figure}[h!]
\includegraphics[width=0.65\textwidth]{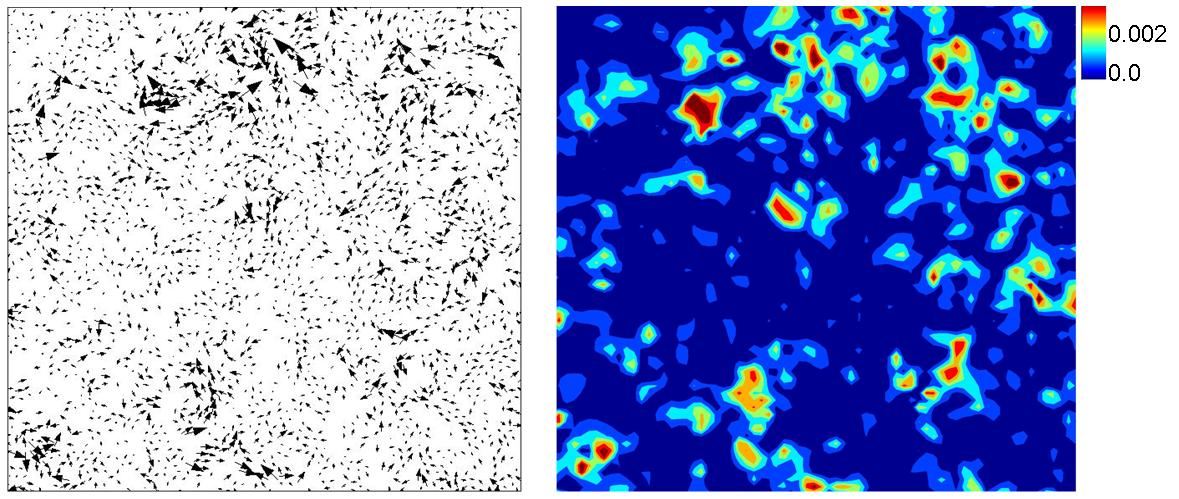}
\caption{Visualization of low frequency modes that are obtained by averaging the $D_{ij}$ over a time interval $\Delta t=500s$.
Left: Example of one of the lowest frequency modes for the present system over the whole field of view $100 \times 100 \mu m^2$. There are regions where the colloidal particles show higher activity than the rest in the field of view. The average participation ratio $P(\omega) = 0.17$.
Right: Contour plot of the local participation ratios for the same mode.}
\label{Eigen-plot}
\end{figure}

Now that we have identified the `active' regions, we will see whether the visualization of the low frequency modes allow us to predict where in the system rearrangements will take place. The existence of these modes was established by doing a normal mode analysis of particle trajectories over the plateau region of MSD \cite{ghosh}-\cite{Kaya}. There are hardly any rearrangements on this time scale, so we can reasonably assume that the particles are close to their local equilibrium, allowing us to determine the normal modes of the system. The displacement of the $i^{th}$ particle from the mean position is given by $u_{i} = x_{i} - <x_{i}>, y_{i} - <y_{i}>$. We then construct the covariance matrix of displacements \cite{ghosh, Kurchan}
%%%%%%%%%%%%%%%%%%%%%%%%%%%%%%%%%%
\begin{equation}
 D_{ij} = \langle u_{i}u_{j} \rangle,
 \label{eq_2}
\end{equation}
%%%%%%%%%%%%%%%%%%%%%%%%%%%%%%%%%%%
where the matrix indices $i,j = 1,2,...2N$ ($N$ particles) runs over both the particle number as well as cartesian components in two dimension. The average $\langle \rangle$ is done over a time interval $\Delta t$. By diagonalizing $D_{ij}$ we obtain $2N$ eigenvalues $\lambda_{i} = 1,2,3..2N$ and their corresponding $2N$ eigenmodes $v(\omega_{i})s$. The frequencies are related to the eigenvalues as $\omega_{i} (= \sqrt{1/\lambda_{i}})$. Note that the above method does not take into account any anharmonicity or damping effects that is present in the system \cite{ghosh}-\cite{Kaya}, \cite{ghosh-P}. The left panel of Fig.~\ref{Eigen-plot} shows the displacement vectors of one of the lowest frequency modes in the system that is obtained by averaging $D_{ij}$ over a time interval $\Delta t=500s$. The quasi-localized nature of the mode is apparent; some smaller regions have larger amplitude than the rest. The degree of localization of these modes can be measured by the average participation ratio following $P(\omega) = [N(\sum_{i} v_{i}(\omega). v_{i}(\omega))^2]^{-1}$, where $v_{i}(\omega)$ is the normal mode (of frequency $\omega$) amplitude projected onto the $i^{th}$ particle. For a strongly localized mode, where the whole amplitude is concentrated over a few particles in the system, $P(\omega)$ scales as $1/N$, whereas for an extended mode it is of order of unity. The average participation ratio of the lowest frequency modes in our system that display quasi-localized character \cite{Harrowell, Laird} lies below $P(\omega) < 0.36$. 

\begin{figure}[h!]
\begin{center}
\begin{minipage}{17.0cm}
\includegraphics[width=0.95\textwidth]{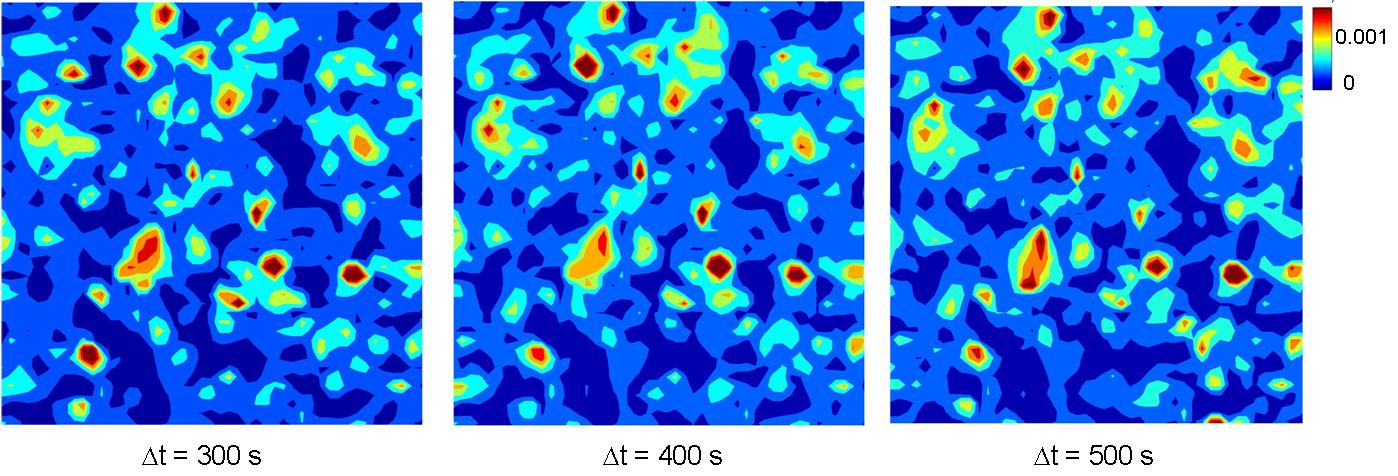}
\end{minipage}
\end{center}
\caption{Contour plots of particle participation ratios, averaged over the lowest $25$ normal modes in a field of view of $60\mu m \times 60\mu m$. This is shown for successively three different time intervals of averaging of $D_{ij}$ : $300s, 400s, 500s$, from left to right. The mode structure remain largely invariant, besides slight variations in few of the quasi-localized zones.}
\label{Eigen-time}
\end{figure}

We will now use the spatial maps of particle participation ratio $p_{i}(\omega) = v_{i}^2$ \cite{Harrowell} to visualize the localized nature of these modes. A contour plot of the particle participation ratios in Fig.~\ref{Eigen-plot} (right) illustrates the spatial structure of the mode shown on the left panel of the same figure. We can use these normal modes to predict the relaxation events provided that their structure remain invariant with time. To test the time invariance, we determine the low frequency modes by averaging $D_{ij}$ over different intervals of time : $300s$, $400s$ and $500s$. Figure~\ref{Eigen-time} shows the contour plots of the particle participation ratios, averaged over the $25$ lowest frequency modes, for different time intervals. We do not observe any significant change in the mode structures, besides slight variations in a few of the quasi-localized zones. This allows us not only to understand the heterogeneous dynamics but also to predict -based on short time measurements of the normal modes- where on long times the rearrangements are likely to happen.

\begin{figure}[h!]
\includegraphics[width=0.6\textwidth]{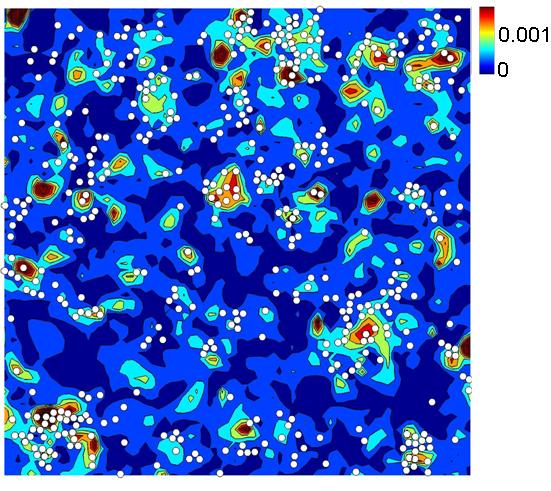}
\caption{Superposition of the particle participation maps, averaged over lowest $25$ modes, and the particles that have relative displacements beyond the cutoff $r_{c}^2$, where $r_{c}=0.245 \mu m$. The normal modes were obtained by averaging the $D_{ij}$ over a time interval $\Delta t=500s$, and the rearranging regions were detected over a time interval $\Delta t=1200s$. The rearranging regions display significant spatial correlation with the quasi-localized zones of the normal modes.}
\label{overlap}
\end{figure}

We therefore compare the contour map of particle participation ratio of the low frequency modes and the rearranging regions to establish a connection between the two. Figure~\ref{overlap} shows a superposition of the contour maps of the particle participation ratio, averaged over $25$ lowest frequency modes, along with the clusters of particles with higher mobility (Fig.~\ref{pdf}) in the field of view of $100 \times 100\mu m^2$. Clearly, the `active' regions where rearrangements happen, exhibit strong spatial correlation with the quasi-localized zones of the low-frequency modes; the rearrangements apparently originate in region of high particle participation ratio. It is worth pointing out that the normal modes were computed by averaging the particle motion over a time interval $\Delta t=500s$, and the rearranging regions were detected in the later stages, over a time interval $\Delta t=1200s$. The spatial distribution of the low frequency modes, thus represents the structure of the system much before the rearrangements occur. The robustness of our results were verified by performing several measurements. For example, see the supplementary figure that shows the contour map of particle participation ratio of the low frequency modes and the rearranging regions at $\phi=0.57$.

In conclusion, we have studied the structural relaxation of a hard-sphere colloidal glass. The dynamics of the particles exhibit spatial heterogeneity : clusters of highly mobile particles are observed. The relaxation of the system occurs through rearrangement of particles in these highly active regions. To understand the nature and origin of these rearranging regions, we have computed the low-frequency modes of the system using normal mode analysis. The low frequency modes display quasi-localized character, which is evident from the contour maps of the particle participation ratios. Our analysis reveals that the regions where rearrangements occur are spatially correlated with the quasi-localized zones of the low frequency modes. This demonstrates that the system indeed relaxes along the softest available modes.

Even though these results are obtained for a colloidal glass under quiescent condition, they should be relevant to weakly driven colloidal glasses \cite{Schall}. It is known for a long time that the plastic deformation in glasses occur in localized regions that are referred as shear transformation zones (STZ) \cite{m_glass,Falk,Schall}. However, their origin remains unclear. Our results should motivate further research along the direction of normal modes to identify the origin of the shear transformation zones.

\textbf{Acknowledgement}: We would like to thank FOM and NWO (vidi grant, P.S) for the financial support of the present research work.

\newpage

%%%%%%%%%%%%%%%%%%%%%%%%%%%%%%%%%%%%%%%%%%%
%%%%%%%%%%%%%%%%%%%%%%%%%%%%%%%%%%%%%%%%%

%%%%%%%%%%%%%%%%%%%%%%%%%%%%%%%%%%%%%%%%%%%
%%%%%%%%%%%%%%%%%%%%%%%%%%%%%%%%%%%%%%%%%

%%%%%%%%%%%%%%%%%%%%%%%%%%%%%%%%%%%%%%%%%%%
%%%%%%%%%%%%%%%%%%%%%%%%%%%%%%%%%%%%%%%%%

%%%%%%%%%%%%%%%%%%%%%%%%%%%%%%%%%%%%%%%%%%%%%
%%%%%%%%%%%%%%%%%%%%%%%%%%%%%%%%%%%%%%%%%%%%%%%%%%%%%

%%%%%%%%%%%%%%%%%%%%%%%%%%%%%%%%%%%%%%%%%%%
%%%%%%%%%%%%%%%%%%%%%%%%%%%%%%%%%%%%%%%%%


\begin{thebibliography}{99}
\bibitem{angell} C.A. Angell, Science 267, 1924-1935, 1995.
\bibitem{c_glass} W. K. Kegel and A. Van Blaaderen, Science 287, 290-293, 2000 ; Y. Gao and M. L. Kilfoil, Phys. Rev. Lett. 99, 078301, 2007.
\bibitem{glass_weeks}E. R. Weeks, J.C. Crocker, A. C. Levitt, A. Schofield, D. A. Weitz, Science 287, 627-631, 2000 ; E. R. Weeks and D. A. Weitz, Phys. Rev. Lett. 89, 095704, 2002.
\bibitem{Schall} P. Schall, D. A. Weitz, F. Spaepen, Science 318, 1895, 2007 ; V. K. Chikkadi, G. Wegdam, D. Bonn, B. Nienhuis, P. Schall, submitted.
\bibitem{shear_weeks} R. Besseling, E. R. Weeks, A. B. Schofield, and W. C. K. Poon, Phys. Rev. Lett. 99, 028301, 2007; D. Chen, D. Semwogerere, J. Sato, V. Breedveld, and E. R. Weeks, Phys. Rev. E 81, 011403, 2010.
\bibitem{ediger} M. D. Ediger, Annu. Rev. Phys. Chem. 51, 99–128, 2000.
\bibitem{Harrowell} A. W. Cooper, H. Perry, P. Harrowell and D. R.  Reichman, Nature Physics 4, 711, 2008; A. W. Cooper, H. Perry, P. Harrowell and D. R. Reichman, J. Chem. Phys. 131, 194508, 2009.
\bibitem{Wyart} C. Brito and M. Wyart, J. Stat. Mech., L08003, 2007; N. Xu, V. Vitelli, A. J. Liu, and S. R. Nagel, Europhys. Lett., 90, 56001, 2010; C. Zhao, K. Tian and N. Xu, Phys. Rev. Lett. 106, 125503, 2011.
\bibitem{ghosh} A. Ghosh, V. K. Chikkadi, P. Schall, J. Kurchan, D. Bonn, Phys. Rev. Lett. 104, 248305, 2010; A. Ghosh \emph{et al.}, Soft Matter 6, 3082-3090, 2010.
\bibitem{Kechen} Ke Chen \emph{et. al}, Phys. Rev. Lett. 105, 025501, 2010.
\bibitem{Kaya}D. Kaya, N. L. Green, C. E. Maloney and M. F. Islam, Science 329, 656-658, 2010.
\bibitem{Kechen1} Ke Chen et. al, http://arxiv.org/abs/1103.2352, 2011. The authors have studied
quasi-2D glasses prepared using microgel colloidal spheres that interact through soft-potential.
\bibitem{hermes_dijkstra} M. Hermes and M. Dijkstra, Euro. Phys. Lett. 89, 38005, 2010.
 \bibitem{Kurchan} R. Mari, F. Krzakala, and J. Kurchan, Phys. Rev. Lett. 103, 025701, 2009.
\bibitem{ghosh-P} A. Ghosh \emph{et. al}, Physica A 390 (18-19), 3061-3068, 2011.
%\bibitem{ghosh1} A. Ghosh, V. K. Chikkadi, D. Bonn, preprint, unpublished.
\bibitem{pusey_megen} P. N. Pusey and W. van Megen, Nature 320, 340, 1986; P. N. Pusey and W. van Megen, Phys. Rev. Lett. 59, 2083, 1987; W. van Megen and P. N. Pusey, Phys. Rev. A 43, 5429-5441, 1991.
\bibitem{SRWilliams} W. van Megen, T. C. Mortensen, S. R. Williams and J. M\"{u}ller Phys. Rev. E 58, 6073-6085, 1998.
\bibitem{Laird}B. B. Laird, H. R. Schober, Phys. Rev. Lett. 66, 636, 1991; B. B. Laird and S. D. Bembenek, J. Phys. Condens. Matter 8, 9569-9573, 1996.
\bibitem{Falk} M. L. Falk and J. S. Langer, Phys. Rev. E 57, 7192, 1998.
\bibitem{m_glass} F. Spaepen, Acta Met. 25, 407-415 (1977); A. S. Argon, Acta Met. 27, 47-58, (1979)


\end{thebibliography}
\end{document}